\documentclass[aoas,MSNbibl,nameyear,dvips]{arximspdf}
\usepackage{graphicx}
\doi{10.1214/13-AOAS614B} %kopijuoti is PTS
\referstodoi{10.1214/12-AOAS614}
\volume{7}
\issue{4}
\pubyear{2013}
\firstpage{1871}
\lastpage{1875}


\begin{document}
\begin{frontmatter}

\title{Discussion of ``Estimating the historical and future
probabilities of large terrorist events'' by~Aaron~Clauset~and~Ryan~Woodard}
\runtitle{Discussion}

\runtitle{Discussion}

\begin{aug}
\author[a]{\fnms{Brian J.} \snm{Reich}\corref{}\ead[label=e1]{reich@stat.ncsu.edu}}
\and
\author[b]{\fnms{Michael D.} \snm{Porter}\ead[label=e2]{mporter@cba.ua.edu}}
\runauthor{B. J. Reich and M. D. Porter}
\affiliation{North Carolina State University and
University of Alabama}
\address[a]{Department of Statistics\\
North Carolina State University\\
Raleigh, North Carolina 27560\\
USA\\
\printead{e1}}

\address[b]{Information Systems, Statistics\\
\quad and Management Science\\
University of Alabama\\
Tuscaloosa, Alabama 35487\\
USA\\
\printead{e2}}

\pdftitle{Discussion of ``Estimating the historical and future
probabilities of large terrorist events'' by Aaron Clauset and Ryan Woodard}
\end{aug}

% HISTORY:
\received{\smonth{7} \syear{2013}}

% ABSTRACT

% KEYWORDS

\end{frontmatter}

We congratulate the authors on this well-written and thought-provoking
paper. They address the problem of estimating the probability of a
large (and rare) terrorist attack by modeling the tail of the attack
size distribution. Recognizing the importance of incorporating
uncertainty, their approach uses bootstrap resampling to obtain a set
of parameter estimates for the tail distribution from which estimates
for the probability of the rare event can be made.
The wide range for the estimated probability of a 9/11-sized attack
(90\% interval $[0.182,0.669]$) illustrates the need to account for
uncertainty in such a problem.

The authors also recognize that the choice of tail model can have a
large impact on the probability estimates. Using multiple tail models
(power law, stretched exponential and log-normal), they estimate that
the probability of a 9/11-sized attack over a 40-year period (or, more
specifically, in 13,274 events) ranges from around 11--35\%. We thought
it would be interesting to compare the results of the authors' analysis
with a more classical extreme value analysis [\citet{deHaan,coles}]
using a generalized Pareto distribution (GPD). The GPD distribution has
three parameters: lower bound $\mu$, scale $\sigma$ and shape $\xi$.
If $Y\sim \operatorname{GPD}(\mu,\sigma,\xi)$, then $Y$'s cumulative density
function is
%
%e1 #&#
\begin{equation}
\label{GPDpdf} % f(y|\mu,\sigma,\xi) = \left[1+\frac{\xi(y-\mu)}{\sigma}\right]^{-(1/
F(y|\mu,\sigma,\xi) = 1- \biggl(1+
\frac{\xi(y-\mu)}{\sigma} \biggr)^{-1/\xi}.
\end{equation}
The shape parameter $\xi$ determines the support of $Y$. If $\xi<0$,
then $Y$ is bounded to the interval $\mu<Y<\mu-\sigma/\xi$; if $\xi
>0$, then $Y$ is unbounded with support $Y>\mu$. The shape parameter
also determines the tail behavior. If $\xi<0.5$, then the density
has\vadjust{\goodbreak}
light tails and finite mean and variance. Large $\xi$ gives heavy
tails. If $\xi>0.5$, the variance in infinite, and if $\xi>1$, then
the mean is also infinite.
If $\xi>0$ and $\sigma=\mu\cdot\xi$, then the GPD reduces to the
(continuous) power-law distribution.

Asymptotic theory suggests that the GPD provides a good approximation
for the tail of a wide range of densities. The typical approach is to
select a lower bound $\mu$ based on exploratory analysis, discard the
data below $\mu$, and estimate $\sigma$ and $\xi$ using maximum
likelihood. %methods and the data above the threshold.
A crucial step in this analysis is to select an appropriate $\mu$
(where $\mu$ is equivalent to the $x_\mathrm{ min}$ used in the article).
If $\mu$ is too small, then the GPD will not fit the tail distribution
and the estimates of $\sigma$ and $\xi$ may be biased. On the other
hand, if $\mu$ is too large, the GPD may fit well, but fewer
observations will be left to estimate $\sigma$ and $\xi$ and their
estimates will suffer from increased variance.

\begin{figure}

\includegraphics{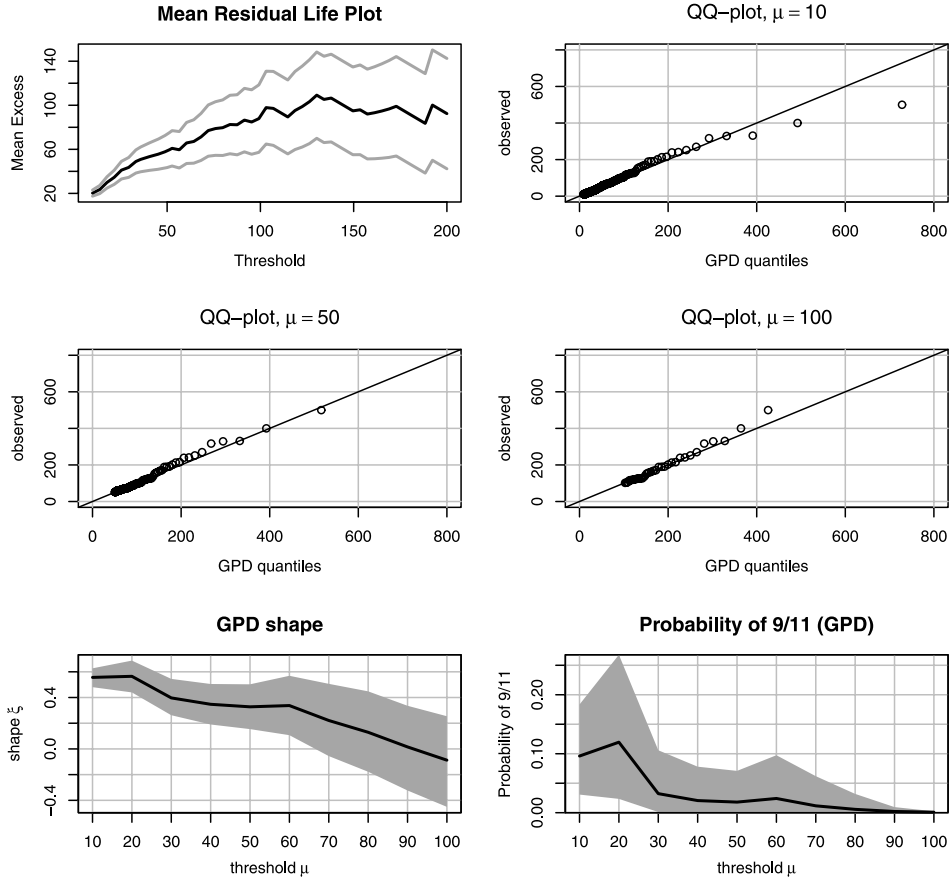}

\caption{Mean residual life plot, qq-plots of the fitted GPD quantiles
versus the quantiles of the data, and the estimated GPD shape and
probability of a 9/11-sized event (shaded region gives bootstrap 90\%
intervals).}\label{f:fig}\vspace*{-3pt}
\end{figure}

A standard exploratory plot [\citet{coles}] used to determine the
threshold $\mu$ is the mean residual life (MRL) plot.\footnote{The analysis uses the \texttt{mrlplot} and \texttt{fitgpd}
functions in the \texttt{POT} package in \texttt{R}. Code is available at
\url{http://www4.stat.ncsu.edu/\textasciitilde reich/Code/}.} Following the authors,
we exclude the 9/11 event and assume stationarity and independence.
Figure~\ref{f:fig} (top left) plots the MRL for the RAND-MIPT
terrorism data.\footnote{\url{http://tuvalu.santafe.edu/\textasciitilde aaronc/rareevents/}.}
If the tail data follow a GPD with a lower bound of $\mu$, then the
MRL plot should be approximately linear for values above~$\mu$.
Therefore, the recommendation is to select the smallest $\mu$ which
gives a linear MRL plot. The authors' selected the threshold for the
power-law distribution by using the value that minimizes the
Kolmogorov--Smirnov statistic between the empirical and fitted
distributions. This approach resulted in a threshold (i.e., $x_\mathrm{
min}$) of around 10. The MRL plot in \ref{f:fig} (top left) suggests
that $\mu=10$ is too small for the GPD, but $\mu=100$ is clearly
sufficient. Below we compare results for thresholds in this range.

%f1 #&#

We fit GPD models, using thresholds spanning 10 to 100, to the original
data (excluding 9/11) and 2000 bootstrap samples. The resulting
qq-plots in Figure~\ref{f:fig} suggest that the GPD with $\mu=10$ may
overestimate the upper quantiles, while $\mu=50$ and $\mu=100$ appear
to provide a better fit. As discussed above, the GPD reduces to the
power law distribution if $\sigma=\mu\cdot\xi$. The bootstrap
intervals of $\sigma-\mu\cdot\xi$ in Table~\ref{t:table} exclude
zero, suggesting the additional flexibility of the GPD model improves
fit for these data.

Figure~\ref{f:fig} (bottom right) also plots the probability of a
9/11-sized event,
%
%e2 #&#
\begin{equation}
\label{f:911} 1- \bigl[{\hat p} + (1-{\hat p})F(y|\mu,{\hat\sigma},{\hat\xi })
\bigr]^n,
\end{equation}
where ${\hat p}$ is the proportion of the events with less than or
equal to $y$ deaths, $F$~is the GPD distribution function, $y= 2749$
is the number of deaths in 9/11, and $n=13\mbox{,}274$ is the number of deadly
terrorism events. For $\mu=10$, the estimated probability is
around\vadjust{\goodbreak}
0.05--0.20, which is similar to the estimate obtained using the
stretched exponential and log-normal methods in the paper. However, the
MRL plot suggests that $\mu=10$ may be too small, and as $\mu$
increases the results change dramatically. The estimate of the shape
parameter $\xi$ decreases from over 0.5 for $\mu=10$ to less than
zero 0 for $\mu=100$. Therefore, when only events above 100 are used,
the estimated density has a light-tail and the probability of a
catastrophic 9/11-sized event decreases to nearly zero.

%t1 #&#
\begin{table}
\caption{Estimates from the GPD model for several threshold values
$\mu$. The 90\% bootstrap confidence intervals are given in the
brackets}\label{t:table}
\begin{tabular*}{\textwidth}{@{\extracolsep{\fill}}lccc@{}}
\hline
\multicolumn{1}{@{}l}{\textbf{Threshold} \textbf{(}$\bolds{\mu}$\textbf{)}} & \textbf{10} & \textbf{50} & \textbf{100} \\
\hline
\# in tail & 853 & 102 & 33 \\
$\Pr(Y>\mu)$ & 0.064 & 0.008 & 0.002 \\
$\xi$ & 0.56 [0.48, 0.63] & 0.34 [0.15, 0.5] & $-0.03$ [$-0.45$, 0.25] \\
$\sigma$ & 9.47 [8.69, 10.29] & 40.98 [31.96, 53.98] & 97.46 [61.54,
152.78] \\
$\sigma-\mu\cdot\xi$ & 3.89 [2.68, 5.25] & 23.82 [8.32, 44.01] &
100.54 [40.65, 193.23] \\
prob. of 9/11 & 0.089 [0.031, 0.183] & 0.01 [0, 0.071] & 0 [0, 0.002]\\
\hline
\end{tabular*}
\end{table}

%f2 #&#
\begin{figure}[b]

\includegraphics{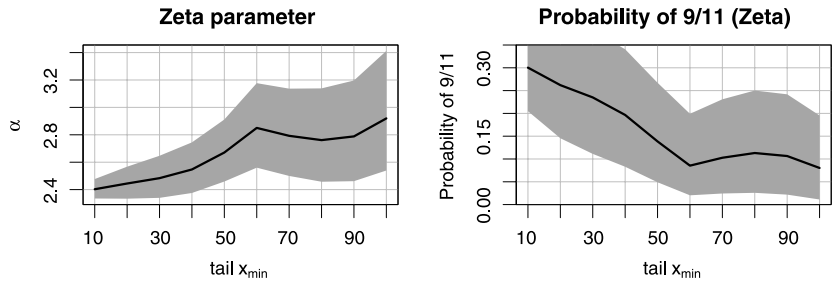}

\caption{The estimated (discrete) power-law parameter $\alpha$ and
probability of a 9/11-sized event
(shaded region gives bootstrap 90\% intervals).}\label{f:fig2}
\end{figure}

Compared to the power-law analysis, this GPD analysis reaches a
different conclusion regarding the likelihood of a 9/11 event. While
the authors' main result is that the likelihood of such a deadly attack
is sufficiently large that it cannot be considered an outlier, the GPD
analysis (following a standard procedure) suggests the opposite. At the
crux of the matter, at least for the GPD analysis, is selecting the
threshold that defines an event as ``extreme.''\vadjust{\goodbreak} Returning to the
bias-variance tradeoff discussed above, the conservative way to resolve
the conflict between results for different thresholds is to use a
larger threshold to reduce bias at the expense of adding variance.

Given the apparent importance of selecting the threshold, it may be
worth discussing this issue from a nonstatistical perspective as well.
One way to motivate an extreme values analysis that discards data below
a threshold is that the processes that govern extreme values are
different than those that govern the bulk of the distribution. For
example, when analyzing extreme precipitation, the bulk of the
distribution results from typical thunderstorms, whereas (at least in
the Southern US) the extreme events are mostly the result of tropical
storms. An analysis which uses data about thunderstorms to infer about
tropical storms is questionable. Returning to the terrorism data,
excluding 9/11, between 1968 and 2008 there were 853 (6.4\%) events
with more than 10 deaths, 102 (0.8\%) events with more than 50 deaths,
and 33 (0.2\%) with more than 100 deaths. Are there really 853 events
that are comparable to 9/11?

While we reach a different conclusion about the probability of a
9/11-sized attack, we do not suggest that the author's analysis is
inappropriate. The results using a power-law distribution (see
Figure~\ref{f:fig2}) are not nearly as sensitive to the choice of
threshold as they are for the GPD. An extreme value analysis is
inherently difficult, and the authors have done a nice job of
justifying their analysis using goodness-of-fit tests, comparing
several models and, perhaps most importantly, making their data and
code available. As stated by \citet{Davison}, extrapolating beyond the
range of data to estimate probability of extreme events ``requires an
act of faith'' in the statistical model and ``a consequence of the lack
of data is that tail inferences tend to be highly uncertain, and that
the uncertainty can increase sharply as one moves further into the
tail.'' We hope this GPD analysis contributes to the discussion about
the sensitivity to model uncertainty.

% imsref loaded by akundreckaite, 2013-10-10 14:01:09
%


%suskaldyti doi

\printaddresses

\end{document}